# The large scale structure of the soft X-ray background. I: Clusters of galaxies

Andrzej M. Sołtan[1,2], Günther Hasinger[3], Roland Egger[1], Steve Snowden[1,4] and Joachim Trümper[1]

[1] Max Planck Institut für extraterrestrische Physik, D-85748 Garching bei München, Germany
[2] Copernicus Astronomical Center, Bartycka 18, 00-716 Warsaw, Poland
[3] Astrophysikalisches Institut, D-14482 Potsdam, Germany
[4] NASA/GSFC, LHEA Code 666, Greenbelt, MD 20771, USA

**Abstract.** Large-scale fluctuations of the X-ray background are investigated using the *ROSAT* All-Sky Survey. The autocorrelation function of the extragalactic background at ∼ 1 keV is determined at separations $0°.1 < \theta < 40°$. We detect a significant signal for $\theta < 6°$, which is larger than previous upper limits. We analyse the cross-correlation between the angular positions of Abell clusters of galaxies and the intensity of the X-ray background. The cross-correlation extends to several degrees for clusters of distance class ≤4. We conclude that Abell clusters are associated with extended, low surface brightness X-ray haloes. The mean luminosity of the diffuse sources in the $0.5 - 2.0$ keV band is $L \approx 2.5 \times 10^{43} h^{-2}$ erg s$^{-1}$ and their characteristic radius $a \approx 10 h^{-1}$ Mpc ($H_0 = 100 h$ km s$^{-1}$ Mpc$^{-1}$). It is shown that the higher density of normal galaxies and AGNs around clusters cannot be responsible for the increased X-ray emission. This is because it would produce a signal in the autocorrelation function measurable by the *Ginga* LAC experiment. Apparent extended emission could be produced either by a number of discrete sources associated with groups of galaxies concentrated around rich clusters or by a hot diffuse gas. It is shown that sources associated with Abell clusters account for ∼ 30 % of the X-ray background fluctuations on scales above ∼ 1°.

## 1. Introduction

The statistical characteristics of the X-ray background (XRB) reflect properties of different classes of objects

Send offprint requests to: A. M. Sołtan, Copernicus Center address

which contribute to this background. At small scales point-like sources produce fluctuations with an angular extent defined by the point spread function of the instrument. Fluctuations at larger scales are created by extended sources as well as by groups of discrete sources. In the 1-2 keV band most (if not all) of the extragalactic XRB is produced by discrete sources (Hasinger et al. 1993). Sources contributing to the XRB constitute a highly heterogeneous group of objects. Practically each class of strong X-ray emitters exhibits its unique characteristics with regard to space and luminosity distributions, spectrum, evolution, clustering properties, etc. This implies that each class of objects produces distinct variations of the XRB and the expected XRB structure is fairly complex.

The problem of background anisotropy has been investigated using several sets of observations by many authors (see Fabian & Barcons 1992 and references therein). The main objective of numerous analyses was to measure the amplitude of the XRB fluctuations and to use it as a constraint on clustering properties of objects creating the XRB (e.g., Persic et al. 1989; Martin-Mirones et al. 1991; Barcons & Fabian 1990; Carrera et al. 1991, 1993; Chen et al. 1994; Sołtan & Hasinger 1994). It was pointed out by Hasinger (1992) and Sołtan & Hasinger, that extended low surface brightness objects can mimic clusters of unresolved sources and any scrupulous study of source clustering should take into account effects produced by extended sources.

In the present paper we investigate the anisotropy of the extragalactic soft XRB on scales between ∼ 0°.5 and 40° using the *ROSAT* All-Sky Survey (RASS) as a data base. This is the first paper of a series which is devoted to the analysis of the large scale effects in the XRB. The objective of the series is to determine relationships between various classes of sources and the soft XRB. In this paper we calculate the autocorrelation function of the XRB and analyse the contribution of Abell clusters to the XRB anisotropy in a wide range of angular scales. The Abell clusters as a class are strong X-ray emitters (Burg et al. 1994 and references therein) and a number of their X-ray properties are established. This makes Abell clusters a suitable sample of objects to be used in the investigation of distinct components of the XRB. Nearby, smaller groups of galaxies constitute another potentially important class of extended X-ray sources which can produce large scale XRB fluctuations and will be investigated in a future paper of this series.

The *ROSAT* All-Sky Survey provides excellent material to study isotropy of the XRB on angular scales above $\sim 10'$ (for the description of the *ROSAT* observatory see Trümper 1983, 1990). The survey combines high angular resolution typical to the imaging X-ray optics with good quality statistical data which cover virtually the whole sky. Excellent performance of the imaging proportional counter PSPC (Pfeffermann et al. 1986) characterized by a very low intrinsic background (Snowden et al. 1992; Plucinsky et al. 1993) allows the examination of extremely low surface brightness features of the XRB in a wide range of angular scales.

The observational data are briefly described below in Sect. 2. The calculations of the autocorrelation function are presented in Sect. 3 and the crosscorrelation of the X-ray data with the Abell catalogue is investigated in Sect. 4. Physical models of the extended emission around clusters are discussed in Sect. 5, and Sect. 6 summarizes the main results of the paper.

## 2. *ROSAT* ALL-Sky Survey

The *ROSAT* All-Sky Survey was performed during a 6-month period following the satellite calibration and verification carried out after launch on 1990 June 1. The RASS exceeds other all-sky surveys of the diffuse X-ray background (e.g., McCammon et al. 1983; Marshall & Clark 1984) in spatial resolution and volume of astrophysically significant data. Due to the scan geometry, the distribution of exposure times in the survey is highly nonuniform. Regions close to the ecliptic poles were scanned more frequently than areas at low ecliptic latitudes. Since the statistical significance of the survey is defined by the photon statistics, the variable exposure introduces strong variations of the signal to noise ratio. For a comprehensive description of the RASS see Snowden & Schmitt (1990) and Voges (1992).

Mapping of the X-ray data in 7 energy bands between 0.1 and 2.0 keV and correction for exposure and non-cosmic background have been carried out using methods described in Snowden et al. (1992, 1994), Snowden & Freyberg (1993), and Plucinsky et al. (1993).

In spite of the careful elimination of different components of the noncosmic background, our final count rate distribution is not completely free from residual contamination. Virtually all X-ray background originating in the satellite environment (mainly scattered solar and auroral X-rays) and instrumental effects ("afterpulse" events) occur in the low and medium *ROSAT* energy bands, typically below 1 keV. The background component produced by high energy charged particles though intrinsically low, becomes noticeable above 2 keV due to diminishing mirror effective area. To minimize the content of foreground, noncosmic photons and contamination by charged particles in our material, we use the hard portion of the *ROSAT* PSPC data excluding, however, energies higher than 2 keV. The data in the gain-corrected pulse-height channels 91 – 131 defined as "R6–band" (Snowden et al. 1994) with the maximum response around 1.15 keV has been selected as this satisfies the above requirements reasonably well. The selection of a relatively narrow range of pulse-height channels increases the uncertainties generated by counting statistics, but allows substantial reduction of systematic effects. One should keep in mind, that the energy resolution of the PSPC is moderate. The energy boundaries of the R6 band defined as those energies where the band response drops to 10 % of its peak value are 0.73 – 1.56 keV (Snowden et al. 1994). Accordingly, from the point of view of the present investigation, the number of soft photons contributing to the R6 band is nonnegligible.

## 3. The autocorrelation function of the XRB

### 3.1. Definitions

Fluctuations of the XRB can be measured using the autocorrelation function (ACF), $w_{XX}(\theta)$. If $\rho_X(\mathbf{n})$ is the intensity of the X-ray background radiation in the direction given by a unit vector $\mathbf{n}$, the ACF is defined as:

$$w_{XX}(\theta) = \frac{\langle \rho_X(\mathbf{n})\rho_X(\mathbf{n}')\rangle}{\langle \rho_X \rangle^2} - 1, \quad (1)$$

where $\langle ... \rangle$ denote the expectation values and $\theta$ is the angle between $\mathbf{n}$ and $\mathbf{n}'$. For the purpose of the present investigation the distribution of the X-ray intensity is represented by an array of count rates in pixels. In the subsequent analysis we use $12 \times 12$ arc min$^2$ pixels. This pixel size, which is roughly consistent with the 90% area of the survey-integrated point spread function, defines the minimum angular scale accessible to our study. As estimator of the ACF we use the expression:

$$W_{XX}(\theta) = \frac{\frac{1}{n_{ij}(\theta)}\sum_{i,j}\rho_X(i)\rho_X(j)}{\overline{\rho_X}^2} - 1, \quad (2)$$

where $\rho_X(i)$ is the count rate in the $i$-th pixel and the sum extends over all pixel pairs with centers separated between $\theta - 6\,\mathrm{arc\,min}$ and $\theta + 6\,\mathrm{arc\,min}$; $n_{ij}(\theta)$ is the number of such pairs in the data and $\overline{\rho_X}$ is the average count rate in all pixels used in the calculations.

*3.2. The ACF of the raw data*

A fundamental obstacle to our study is posed by thermal emission of hot plasma within the Galaxy (McCammon & Sanders 1990; Hasinger 1992). We are interested in the fine variations of the extragalactic component of the XRB, while a substantial fraction of the All-Sky data is strongly dominated by Galactic emission. At low galactic latitudes large areas showing higher absorption by cold gas are also clearly visible (Snowden et al. 1995).

Apart from general diffuse galactic emission, many discrete features are present in the All-Sky maps. In the south galactic hemisphere several areas of enhanced surface brightness including the Magellanic Clouds and the Eridanus region make it difficult to select a coherent and large enough section of the sky suitable for investigation of the large scale structure of the XRB. In the northern hemisphere, the situation is less complex. A large area at galactic longitudes between 260° and 50° (through the Galactic center) is dominated by the North Polar Spur and the Virgo Cluster, but the remaining region is apparently free from obvious galactic and nearby extragalactic contamination. Fig. 1 shows a section of the RASS around the North Galactic Pole (NGP). The data are displayed in the polar equal area projection in galactic coordinates centered at the NGP. The half-circle contour has radius of 50° and depicts the area used in the analysis (see below). To visualize large scale features, the original high resolution data have been smoothed in Fig. 1 using median filter with radius of 3°. Bright areas in the lower half of the figure show saturated images of the North Polar Spur and the Virgo cluster.

Judging from the visual appearance of X-ray maps, we have selected the area on the northern hemisphere defined in galactic coordinates as: $70° < l < 250°$ and $b > 40°$ (see Fig. 1) for a further study. Two objects, the Coma Cluster and Abell 1367 are particularly conspicuous X-ray sources in that region and produce count rates distinctly larger than any other source. To avoid the situation that our results are dominated by just two objects we have removed areas around these clusters from further analysis. Eventually, our data contain X-ray count rates in more than 91000 pixels and cover an area of $\sim 3670\,\mathrm{deg}^2$. The average R6 count rate in the area $\overline{\rho_X} = 79 \times 10^{-6}$ PSPC cnt s$^{-1}$ arc min$^{-2}$ with an rms of $79 \times 10^{-6}$ in the same units. The high rms is produced by several hundreds of strong discrete sources and these sources contribute substantially to the noise in our measurements of the ACF. Count rates in 584 pixels exceed $300 \times 10^{-6}$ ($\approx \overline{\rho_X}+3\cdot\mathrm{rms}$). To reduce the impact of a relatively small number of pixels with the highest count rates we have reduced the count rates above $300 \times 10^{-6}$ to this value. The background subtraction introduces additional relative scatter into the data; this is illustrated, for instance, by the fact that the formal net count rate in 1279 pixels in the area is negative. For the same reason as for the high count rate pixels, we have set the count rate to $10 \times 10^{-6}$ in all pixels with registered count rates below this limit. Introduction of the low and high count rate thresholds affects only about 4.4 % of all pixels but reduces the rms of the count distribution almost by a factor of 2 down to $47 \times 10^{-6}$ in the above units.

**Fig. 1.** Section of the RASS around the North Galactic Pole in equal area projection. Data are smoothed using a median filter with radius of 3°. The half-circle with radius of 50° defines the area used to determine the ACF and the correlations with Abell clusters. The lower part of the figure shows the highly overexposed North Polar Spur and the Virgo cluster.

The ACFs of the XRB in bands R5 and R6 in the selected region are shown in Fig. 2a. Details of the ACF in R6 at small separations are also shown in Fig. 2b. Error bars plotted in Fig. 2b for some points denote 1 $\sigma$ scatter calculated under the assumption that the data are perfectly uncorrelated. These uncertainties represent only the noise due to count rate fluctuations in individual pixels.

We have also calculated the ACF for areas extending outside the above limits to estimate the effects of the North Polar Spur and/or galactic plane. It appears that small border changes do not affect the ACF noticeably, while larger ones (e.g., $50° < l < 270°$) produce a substantial increase of the ACF amplitude at all calculated separations.

*3.3. The ACF of the extragalactic XRB*

Plasma emission dominates mostly at soft X-ray energies (Hasinger 1992). To estimate and subtract this galactic contribution to the fluctuations in the band R6, we utilize data in channels 70 − 90, which define the softer, R5 band (Snowden et al. 1993) centered at 0.83 keV and extending approximately between 0.56 keV and 1.21 keV.

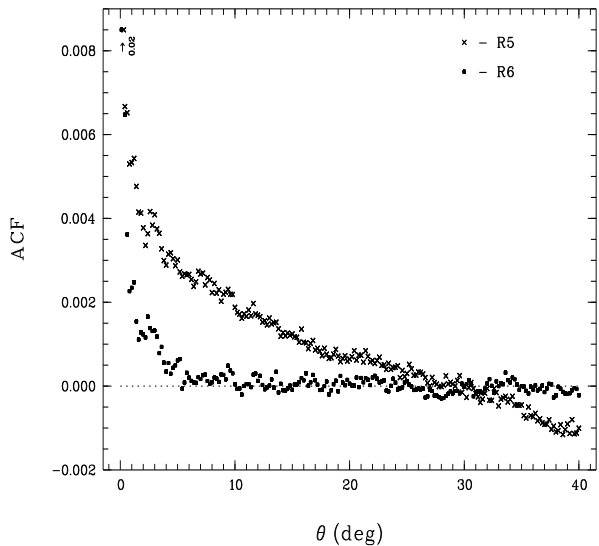

**Fig. 2a.** The ACF of the RASS band R5 and R6.

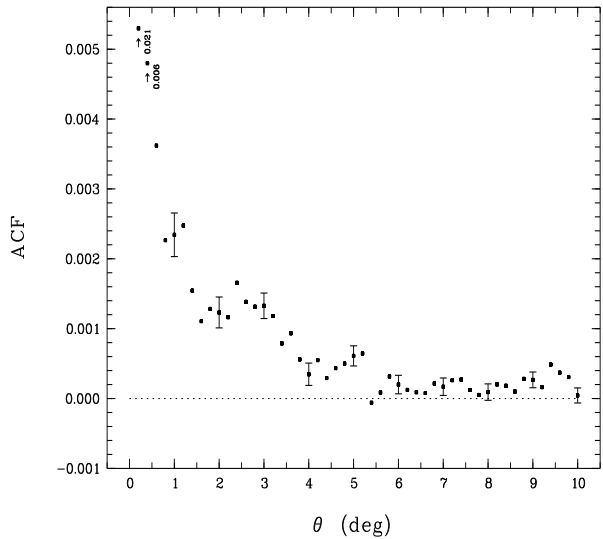

**Fig. 2b.** The ACF for $\theta < 10°$ of the RASS band R6. The error bars shown for some points represent the rms scatter calculated for strictly uncorrelated data.

The ACFs of the R5 and R6 data sets are distinctly different. This implies that fluctuations of the XRB in those two bands are at least partially independent. Moreover, the conjecture that a *significant* fraction of the ACF amplitude in R6 data results from the contamination by soft photons is unlikely. This is because the strong ACF signal in the R5 data extends to large separations, where the R6 count rate distribution shows no evidence for significant fluctuations. Assuming that most of the ACF signal in R6 comes from the low energy flux, one should expect a proportional behaviour of both ACFs on all scales. This is not observed. For instance, the R5 ACF amplitudes at separations $2° - 3°$ and $10° - 14°$ are $3.8 \times 10^{-3}$ and $1.6 \times 10^{-3}$, respectively. For the same separations the ACF of the R6 count rates amounts to $1.4 \times 10^{-3}$ and $5 \times 10^{-5}$, and the latter amplitude is consistent with zero. It shows that the decay of the correlation is at least an order of magnitude steeper in the R6 than in R5 data, and points out that the underlying fluctuations have different origin.

With the exception of the smallest separations ($\theta < 0°.5$), the ACF amplitude in the R5 band lies significantly above that in R6. It seems natural to assume that the ACF of the R5 band measures mainly the fluctuations produced by soft Galactic emission. One could also expect that some, non-dominating fraction (as noted above) of the ACF signal in R6 is produced by the same mechanism, because of penetration of the soft photons into this band. To separate galactic and extragalactic signals in R6 we make specific assumptions about the spectral characteristics of the fluctuations in these two XRB components. Let $S$ and $H$ denote the count rates in band R5 and R6, respectively. The total rates are sum of the galactic ($g$) and the extragalactic ($e$) flux:

$$\begin{aligned} S &= S_g + S_e \\ H &= H_g + H_e. \end{aligned} \qquad (3)$$

It is now assumed that both components satisfy the relationship:

$$\begin{aligned} S_g &= k_g H_g \\ S_e &= k_e H_e, \end{aligned} \qquad (4)$$

where $k_g$ and $k_e$ are constant. Eq.(4) postulates that fluctuating parts of both XRB components have the same spectral slopes as their respective average intensities. Since the galactic component is softer than the extragalactic one, $k_g > k_e$. Using Eqs. (3) and (4) the extragalactic count rates in band R6 are given by:

$$H_e = \frac{k_g H - S}{k_g - k_e}. \qquad (5)$$

The spectrum of the extragalactic XRB is well approximated by a power law with the energy spectral index of $\sim -1$ (Hasinger 1992). Using the calibration of the *ROSAT* telescope/PSPC combination we find that this slope yields $k_e = 0.66$. (For an energy spectral index of -0.7 this value would be 0.61). The average hydrogen column density in the area is $N_H = 1.8 \times 10^{20} \mathrm{cm}^{-2}$ and this value was assumed to calculate low energy absorption by cold gas, but the results are only weakly sensitive to $N_H$. The value of $k_g$ is not directly estimated from observations. Our model effectively assumes that the whole fluctuating component of the R6 data which is softer than the extragalactic XRB originates in the Galaxy. Thus, the "best estimate" of the parameter $k_g$ is determined by minimizing the ACF of the $H_e$ distribution defined by Eq. (5). We have calculated the ACF for several values of $k_g$. The minimum amplitude of the ACF has been obtained for $k_g \approx 5$. This value corresponds to the temperature of $1.0 \times 10^6$ K for thermal bremsstrahlung radiation and is in good agreement with measurements by Hasinger (1992). The ACF of the data "rectified" in this way is very similar to the uncorrected R6 ACF and is shown in Fig. 2c with dots (the curve defined by crosses is discussed in Sect. 5.4 below).

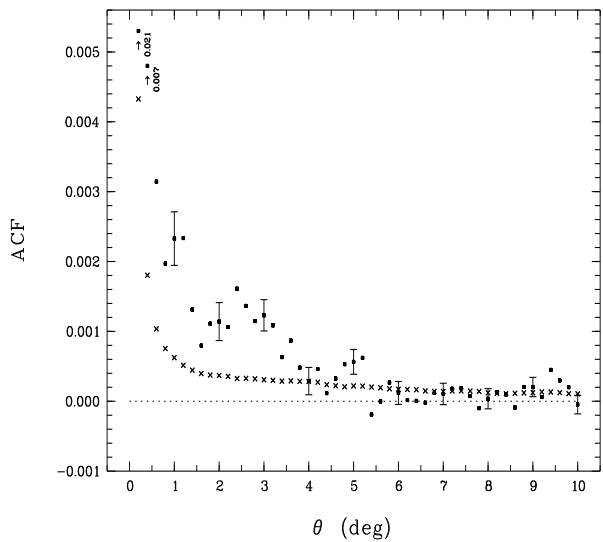

**Fig. 2c.** The ACF of the extragalactic component of the XRB defined by eq. (5) – dots, and the predicted ACF produced by Abell clusters (Sect. 5.4) – crosses. The ACF due to the Abell clusters is calculated assuming that X-ray sources associated with clusters are described by a model given in Sect. 4.4 and using the actual distribution of clusters.

In addition to the narrow peak for $\theta < 0°.7$ (see below), the original ACF in the R6 band shown in Figs. 2a and 2b exhibits a clear positive signal below $\sim 5°$ with an average of $8.5 \times 10^{-4}$ for $1°.3 < \theta < 5°$. Between $6°$ and $10°$ the amplitude of the ACF is still positive and oscillates around $2.0 \times 10^{-4}$. After rectifying the R6 distribution according to Eq. (5) the amplitude of the ACF is reduced by a similar amount in both separation ranges. Between $1°.3$ and $5°$ the ACF $W_{XX} = (7.5 \pm 0.6) \times 10^{-4}$ and between $6°$ and $10°$ it is now $(1.1 \pm 0.3) \times 10^{-4}$. The errors represent only the statistical noise. We are aware of some shortcomings of our method to isolate extragalactic fluctuations. In particular, if fluctuations of the XRB are generated by sources with different spectra than that of the average background, the present procedure would produce incorrect results. Nevertheless, it is clear that fluctuations in the R6 band are of a different nature than those in R5 and that most of fluctuations detected in R6 could not be explained by contamination from soft galactic emission. To substantiate this statement we have calculated the cross-correlation functions (CCF) between the R5 and R6 raw and rectified data. At separations larger than the extent of the point spread function, a positive signal in the R5 - R6(raw) CCF results from fluctuations of galactic and extragalactic origin while that for the R5 - R6 rectified CCF the signal comes only from the extragalactic fluctuations. This leads to a significant reduction of the CCF amplitude at all separations. Between $1°.3$ and $5°$ the average CCF amplitudes are equal to $1.3 \times 10^{-3}$ and $7.8 \times 10^{-4}$ for raw and rectified R6 band, respectively, and between $6°$ and $10°$ the we obtain $6.5 \times 10^{-4}$ and $2.2 \times 10^{-4}$. One should note that the galactic flux, which produces most of the fluctuations in the R5 band, does not dominate the total count rate in this band and that variations of the extragalactic components generate fluctuations of comparable relative amplitude in the R5 and R6 band. Using the amplitudes of the CCF and the ACFs in these two bands we have estimated the galactic contribution to the fluctuations in the R6 rectified data. At all separations larger than $\sim 0°.5$ the amplitude of the ACF due to the soft contamination is consistent with zero within statistical errors. Between $1°.3$ and $5°$ it is below $0.6 \times 10^{-4}$ and between $6°$ and $10°$ - below $0.3 \times 10^{-4}$. We conclude that the ACF shown in Fig. 2c adequately describes the XRB fluctuations at energies $0.7 - 1.6$ keV and that these fluctuations are most likely of extragalactic origin.

The present data cannot be used to determine the ACF at separations smaller than the pixel size (12'), i.e., for $i = j$ in Eq. 2. This is because fluctuations of the count rates are dominated by the discrete nature of photon statistics precluding estimates of the ACF if pixels overlap. Random noise which produces pixel to pixel fluctuations is usually described by Poisson statistics. This allows the subtraction of a "Poisson term" from the sum at the left hand side of Eq. (2) and thus to calculate $W_{XX}$ averaged over separations corresponding to "$i = j$" (see for instance Peebles 1980). In our case, however, this method could not be applied. The rectified R6 data used to calculate the ACF are determined using a multi step procedure in which several background components are subtracted. Because of the varying level of the background rate and nonuniform distribution of exposure times over the area, the final uncertainty of the count rate ascribed to a single pixel depends strongly on the pixel position. In effect, the fluctuations of the net count rate are not described by simple Poisson statistics.

The positive amplitude of the ACF for separations up to $\sim 5°$ is intriguing. Its reality is unquestionable, but it is difficult to assess total uncertainties involved in the ACF measurements. Both visual examination of the *ROSAT* maps and the apparent signal in the ACF show that structures with a size of several degrees are present in the XRB. Some insight into the character of these structures could be obtained by splitting the area into several smaller regions and calculating the ACF separately for each segment. We have divided the original area into 7 pieces of similar size: one polar region with $b > 70°$ and six with $40° < b < 70°$ each extending over $30°$ in galactic longitude. The ACF in these areas reveals some differences. The minimum and maximum ACF amplitudes averaged between $0°\!\!.7$ and $3°\!\!.1$ are $-0.1 \times 10^{-3}$ and $2.8 \times 10^{-3}$, while for the whole area $W_{XX}(0°\!\!.7 - 3°\!\!.1) = 1.2 \times 10^{-3}$. This indicates that there are noticeable field to field variations of the fluctuation level, since the rms scatter expected just due to statistical fluctuations for the uncorrelated data is of the order of $0.2 \times 10^{-3}$. However, in 5 of 7 areas the ACF amplitude exceeds $0.6 \times 10^{-3}$, indicating that the fluctuations of measurable amplitude are widespread over most of our data and the average ACF signal is basically representative for the entire XRB distribution.

A weak but persistent ACF signal of $1.1 \times 10^{-4}$ is present in the data between $5°$ and $10°$, although we could not rule out the possibility that it results from some imperfections in our procedure to remove the soft fluctuating component from the R6 band. Our 7 areas are too small to get significant estimates of the ACF at such large separations. The statistical uncertainty of the ACF amplitude averaged over scales $5°$ to $10°$ is equal to $0.3 \times 10^{-4}$.

### 3.4. Comparison with previous work

To compare our measurement with several previous results obtained with *ROSAT* as well as with other experiments, we plot the relevant data in Fig. 2d. The present results are shown with dots. Error bars represent $2\sigma$ uncertainties expected for uncorrelated data. A short dotted line at $\theta < 0°\!\!.3$ represents the estimate of the ACF determined by Sołtan & Hasinger (1994). Using several dozen *ROSAT* pointed observations, they obtained $2\sigma$ upper limits at a level of $\sim 2 \times 10^{-3}$ on scales larger than $3'$. One should note that this value is not in conflict with the large ACF amplitude below $0°\!\!.5$ obtained in the present investigation. Firstly, the method applied by Sołtan & Hasinger explicitly eliminated from the ACF any contribution by fluctuations with a characteristic length exceeding the size of a single field i.e., $\sim 30'$. This effectively reduced the ACF amplitude on all smaller scales. Secondly, the present ACF describes the total XRB fluctuation, including those produced by point and extended sources, while the Sołtan & Hasinger investigation concentrated on fluctuations generated by the non-random distribution of weak point-like sources, and all known extended sources have been deliberately removed from their data. The low amplitude of the ACF produced by small angular scale fluctuations found in that work puts strong constraints on the level of clustering of sources contributing to the XRB. It indicates that the positive signal at large scales reported in the present paper does not result from the source clustering but is produced mostly by extended sources.

The dashed line approximately shows 95 % upper limits obtained by Chen et al. (1994) in the range of $8' - 300'$ using 1 deep and 6 medium exposure *ROSAT* fields which span $2° \times 7°$ of the sky. At separations above $\sim 1°$ our detections are marginally consistent with the Chen et al. upper limits but there is a significant discrepancy between both investigations at smaller separations. The reason for this apparent disagreement is exactly the same as for results by Sołtan & Hasinger: at separations below 0.5 degrees our data include a relatively strong contribution by the core emission of clusters of galaxies (see below) and other bright sources, while the Chen et al. data have been deliberately chosen in a source-free region.

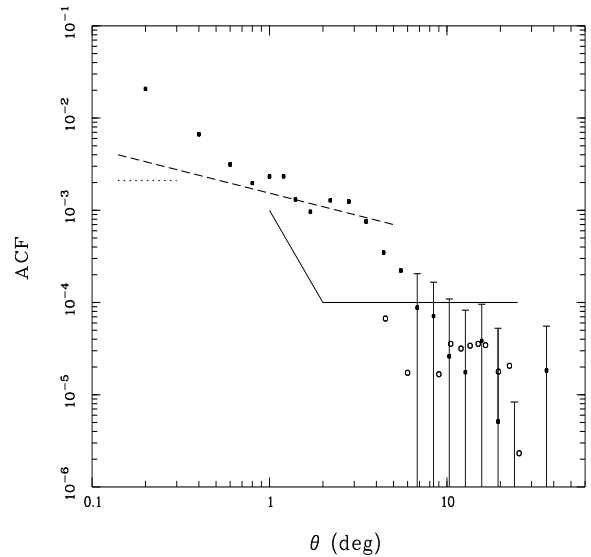

**Fig. 2d.** Selected results on the ACF of the XRB. Full dots – present results ($2\sigma$ error bars are shown for data points with $S/N < 2$); dotted line – $2\sigma$ upper limit determined by Sołtan & Hasinger (1994) using *ROSAT* pointings excluding known extended sources; dashed line – 95 % upper limits derived by Chen et al. (1994) from $2° \times 7°$ area observed with *ROSAT*; solid line - 95 % upper limits found by Carrera et al. (1991, 1993) using *Ginga* LAC 4 – 12 keV observations; open circles – Mushotzky & Jahoda (1992) measurements obtained with the *HEAO-1* A-2 experiment (2 - 10 keV).

Carrera et al. (1991, 1993) measured the ACF of the XRB in the $4 - 12$ keV energy band using *Ginga* LAC observations. Due to the relatively large solid angle of the point response function of the *Ginga* data ($\sim 1° \times 2°$) the effect of point sources had to be carefully modelled and subtracted from the observed ACF. Below $1°$ the uncertainties entangled in this procedure are high and the data are probably not reliable. However, at separations above $1°$ Carrera et al. give stringent upper limits for the ACF. Their results – shown schematically in Fig. 2d with solid lines – can be summarized as follows: at separations $2° - 25°$ the ACF does not exceed $\sim 10^{-4}$ at 95 % confidence level and below $2°$ the upper limit increases steadily and reaches $\sim 10^{-3}$ at $1°$. These figures are in satisfactory agreement with our ACF measurements only above $5°$. Between $5°$ and $10°$ a possible signal in the ROSAT data of $\sim 1 \times 10^{-4}$ is marginally compatible with the *Ginga* upper limits, and there is also no clear discrepancy above 10°. Our measurements averaged between 10° and 40° give $W_{XX}(10°-40°) = -4 \times 10^{-5}$ with a scatter of $12 \times 10^{-5}$ and no systematic trends exceeding the *Ginga* limits are visible. Below 5° the amplitude of the ACF determined using the *ROSAT* data runs systematically above the *Ginga* upper limits. This discrepancy indicates that the fluctuations of the XRB depend strongly on energy. The *ROSAT* and *Ginga* experiments cover well–separated energy bands, and fluctuation analyses in those bands are likely to provide information on different features of the XRB.

Open circles in Fig. 2d show results obtained by Muschotzky & Jahoda (1992) using the *HEAO-1* A-2 data (2 − 10 keV). They report the ACF detection at scales of 6°−20° of $\sim 3 \times 10^{-5}$. Recently Jahoda (1993) reevaluated *HEAO-1* observations and noted that the whole ACF signal is generated in low supergalactic latitudes ($|B| < 15°$). Our measurements, give formally $W(\theta) \sim (2.5\pm1.4) \times 10^{-5}$ (the error represents statistical fluctuations only). This is in agreement with the *HEAO-1* A-2 results since the region investigated in the present paper also lies at low supergalactic latitudes and more than half of our data are from $|B| < 15°$.

The positive amplitude of the ACF found in the present investigation for a wide range of separations poses a question of possible sources producing large scale fluctuations in the soft extragalactic XRB. To examine this problem we intend to determine the correlations of known classes of objects with the XRB and we begin our study with rich clusters of galaxies.

## 4. Cross-correlation of X-ray maps with the Abell catalogue

In this investigation we concentrate on the large angular scale effects produced in the XRB by Abell clusters. Abell clusters constitute a characteristic class of objects with well known optical and X-ray properties. Analogous effects generated by smaller groups and clusters will be analysed in a separate paper.

For the purpose of the present study all clusters in the sky region under consideration were divided into three samples according to their distance classes (DC): I − DC = 1 to 4, II − DC = 5, and III − DC = 6. Samples II and III contain clusters with richness class RC $\geq 1$. Sample I − due to the small number of objects − includes also clusters with RC = 0. After exclusion of Coma and A 1367 (see above), samples I, II and III contain 56, 210 and 312 clusters, respectively.

### 4.1. Method of estimation

The cross-correlation function (CCF) is defined in a similar way as the ACF:

$$w_{Xc}(\theta) = \frac{\langle \rho_X(\mathbf{n}) \rho_c(\mathbf{n}') \rangle}{\langle \rho_X \rangle \langle \rho_c \rangle} - 1, \quad (6)$$

where $\rho_c(\mathbf{n})$ is the surface density of clusters in the direction $\mathbf{n}$. We estimate $w_{Xc}(\theta)$ as follows. The cluster distribution is binned into pixels exactly the same as the pixels used for the X-ray maps. Thus, $\rho_c(i)$ is equal to 0, 1, 2,.. according to the number of clusters found in the $i$-th pixel. Then, the expectation values in eq. (6) are substituted by their respective averages obtained from the data:

$$W_{Xc}(\theta) = \frac{\frac{1}{n_{ij}(\theta)} \sum_{ij} \rho_X(i) \rho_c(j)}{\overline{\rho_X}\;\overline{\rho_c}} - 1. \quad (7)$$

The CCF describes the coupling between the cluster distribution and the intensity of the X-ray background radiation and for randomly distributed clusters it defines the average contribution of a single cluster to the XRB.

### 4.2. Correlation of the XRB with nearby clusters

The CCF of the rectified R6 count rate distribution and cluster sample I has been calculated according to Eq. (7) and the results are shown in Fig. 3. The prominent peak at separations below 0°.5 (first 3 data points are shown not to scale) demonstrates the established fact that Abell clusters as a class are bright X-ray emitters. Apart from this, there is a positive signal at a level of $\sim 0.01$ for $\theta < 7°$. The statistical significance of this effect deserves a careful study and we discuss it below. At all larger separations the ACF oscillates around zero with a typical amplitude substantially smaller than the apparent signal at small $\theta$.

Uncertainties of the CCF measurements are difficult to assess. The positive amplitude of the CCF indicates the existence of a systematic excess of the X-ray intensity around clusters in sample I. As a first approximation we may assume that the relative enhancement of the XRB around clusters is on the average equal to the CCF: $\delta\rho_X/\rho_X = W_{XX}(\theta < 7°)$. Fluctuations of the XRB with a typical amplitude of 0.01 would produce an ACF signal of $\sim 10^{-4}$. This is comparable with the actually observed ACF on scales larger than 5° but falls short by an order of magnitude at separations below $\sim 4°$. The XRB fluctuations at these separations estimated from the ACF are approximately equal to $\delta\rho_X/\rho_X \approx 0.03$. The existence of fluctuations with an amplitude 3 times larger than that implied by the CCF, i.e., fluctuations apparently not related to the clusters in sample I, substantially increases the uncertainty of the CCF estimates. Moreover, clusters are distributed non-randomly within the area and create several clumps. It is conceivable that a random coincidence of a single positive XRB fluctuation with a group of clusters could produce the signal visible in Fig. 3. The uncertainties of the CCF are described by the number and amplitude of peaks and depressions in the XRB and by the number of cluster clumps, rather than the rms scatter of the XRB and the total number of clusters. The combined effects of those factors on the total uncertainty of the CCF could be assessed only by means of an appropriate simulation scheme. Typical CCFs expected for uncorrelated data and associated uncertainties have been calculated using the original X-ray maps and simulated cluster distributions. "Random" cluster samples were created by rotation of the map of the Abell clusters around the galactic polar axis. This procedure preserves the statistical characteristics of the data and therefore provides realistic estimates of total uncertainties. To generate a set of CCFs which are statistically independent at separations of several degrees one needs large rotation angles. Unfortunately this requirement limits the maximum number of independent simulations. Interesting separations in the sample I extend to 7° and we chose 45° as the rotation angle. As a result we got just 7 simulated CCFs. It is doubtful that such a small number is sufficient to provide accurate error estimate but we believe that it can still give a fair understanding of uncertainties involved in the CCF estimates.

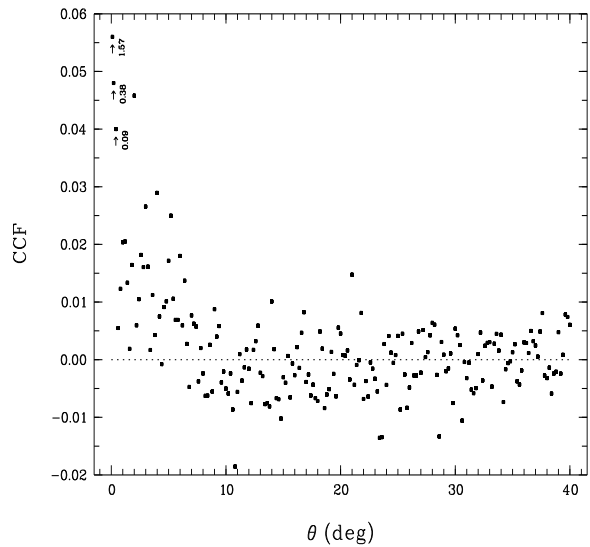

**Fig. 3.** The CCF of the XRB and clusters in sample I. The first three points are shown not to scale. A positive signal at separations below ∼ 7° indicates some enhancement of the XRB surface brightness around clusters.

The rms scatter among the simulations is taken as the $1\sigma$ uncertainty of the CCF. The average amplitude of the CCF for separations $1° < \theta < 6°$ is $1.32 \times 10^{-2}$ while the average amplitude and the rms scatter in 7 simulations amount to $0.14 \times 10^{-2}$ and $0.41 \times 10^{-2}$, respectively. The maximum amplitude obtained in the simulations is equal to $0.79 \times 10^{-2}$ (= 60 % of the real signal). One should expect that the average simulated CCF is equal to zero and our results are consistent with this value. Thus, assuming that the "true" average simulated CCF $\overline{W_{Xc}^{sim}} \equiv 0$, the signal to noise ratio of the present detection $S/N = 3.3$. On the other hand, rotation of the original cluster data around the galactic polar axis does not remove a possible correlation with galactic latitude. If the XRB distribution also shows a similar correlation, the positive amplitude of the average simulated data is conceivable. Thus, to estimate the significance level of our detection, it is necessary to subtract the average simulated signal of $0.14 \times 10^{-2}$ from the observations. In this case $S/N = 2.9$. Because the "noise" in the denominator is calculated using a sample of 7 simulations only, the $S/N$ ratio does not constitute a precise criterion to establish the statistical significance of the signal detection. Using one-sided tolerance limit statistics for a normal distribution, we may state that simulated CCFs would produce by chance the signal stronger than that actually observed in less than $P = 2$ % of trials.

### 4.3. Correlation of the XRB with DC = 5 and 6 clusters

If the weak extended signal detected for nearby clusters is real, it should be visible also in more distant clusters. Obviously, DC = 5 clusters are expected to produce a weaker signal than DC ≤ 4 ones, but the sample II contains almost 4 times more objects than sample I and the uncertainties of the CCF measurements are substantially reduced. The CCF between the sample II distribution and the XRB is shown in Fig. 4. The calculations have been performed in the same way as for the sample I. The narrow peak for $\theta < 0°.3$ is very prominent as well as a signal with the amplitude of a few hundreths extending to ∼ 3°. Correlations on scales < 3° contrast with the apparent lack of signal on larger scales. In particular, the CCF averaged over 5° to 10° is equal $(6.7 \pm 6.6) \times 10^{-4}$, where the $1\sigma$ uncertainty of the mean represents the statistical noise only. At all larger separations the CCF also does not deviate significantly from zero. Similar to sample I, total uncertainties of the CCF amplitude below 3° have been estimated by means of simulations. The smaller angular extension of correlations in the sample II as compared to sample I allows us to create larger number of randomized CCFs. We generated 11 simulated cluster distributions with a rotation angle of 30°. The CCF averaged over separations $0°.7 - 3°$ is equal to $1.81 \times 10^{-2}$ while the simulations give an average and rms of $0.31 \times 10^{-2}$ and $0.68 \times 10^{-2}$, respectively. The signal to noise ratio reaches 2.5 assuming that $\overline{W_{Xc}^{sim}} \equiv 0$ and 2.2 with $\overline{W_{Xc}^{sim}} = 0.31 \times 10^{-2}$. Similarly to the sample I results, we find that simulated CCFs produce by chance a signal stronger than that actually observed in less than $P = 3.5$ % trials. The narrow separation range in which the CCF signal is strong in the distant clusters as compared to the nearby ones generates a larger scatter between simulations in the sample II and effectively reduces significance of the signal detection.

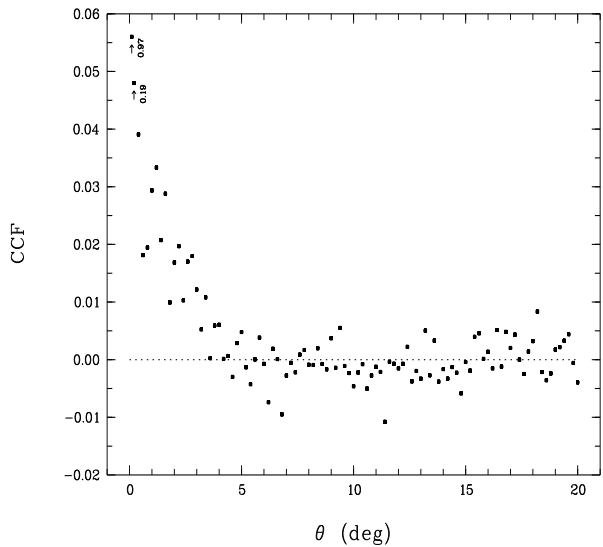

**Fig. 4.** Same as Fig. 3 for sample II.

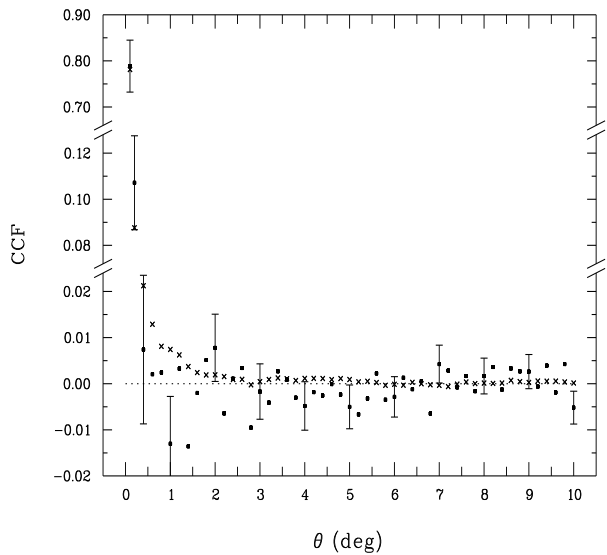

**Fig. 5.** The CCF of the XRB and clusters in sample III – dots, and the single component model CCF (Sect. 4.4) – crosses.

Statistical characteristics of the CCFs of sample I and II are quite different but assuming that both samples are independent, the probability to generate signals in simulations at least as strong as in the real data in both samples simultaneously is $6.7 \times 10^{-4}$.

Sample III has been analysed in a similar way as the sample II. The resultant CCF is shown in Fig. 5 with dots. Error bars are $1\,\sigma$ uncertainties due to count rate fluctuations in individual pixels. The total uncertainties determined from simulations are typically larger by 50 %. Apart from a peak at separations below $0°\!.3$ the data do not exhibit distinct features at larger separations. In sample II correlations on scales of $3°$ are clearly visible. Because the average redshift of clusters in the sample III is about 50 % larger than the redshift in sample II, we expected that correlations in sample III would extend to $\sim 2°$. At first glance the data are inconsistent with these predictions. The CCF averaged between $0°\!.3$ and $2°$ is equal to $-0.01 \times 10^{-3}$ and the rms scatter among 11 simulations amounts to $6.0 \times 10^{-3}$. In the next section this question is analysed quantitatively, and it is shown that the apparent lack of correlations at $\sim 1°$ is statistically insignificant.

It is shown in the next section that standard cluster emission described by $\beta$ model with a core radius $a_0 = 0.125\,h^{-1}$ Mpc (e.g., Henry et al. 1992) is represented in our CCFs by a sharp peak below $\theta \approx 0°\!.5$. Since two independent cluster samples exhibit fairly strong positive correlations with the XRB on angular scales an order of magnitude larger, some new type of relationship between clusters and the XRB is strongly suggested. Assuming that the CCF signal is generated by a physical association of clusters and the XRB, one can expect that the CCF should scale with distance. Below we investigate this question using our three cluster samples.

### 4.4. A model for the CCF

Using a simple model we intend to evaluate typical sizes and luminosities of sources which generate correlations between the cluster distribution and the XRB. The wide range of distances to clusters and the broad X-ray luminosity function imply a significant variation of the angular extents and fluxes of sources in our samples. On the other hand, the CCF amplitude measures the contribution to the XRB of a randomly chosen cluster, i.e., it defines the average X-ray properties of sources associated with clusters. To model the observed CCF we assume, that all clusters in each sample separately have similar X-ray surface brightness distributions. To illustrate how serious simplifications are introduced by this assumption we note that within each sample the rms of the redshift distribution is roughly equal to 30 % of the mean sample redshift. This signifies that both source diameters and fluxes vary substantially. Nevertheless, this simple model is capable of providing correct estimates of the average cluster luminosities and to account for the small scale CCF features (see below). According to this assumption, the source brightness distribution is fixed as:

$$S(\theta) = S_0 \cdot f(\theta), \qquad (8)$$

where $S_0$ is the total flux (the same for all clusters in a sample) and $f(\theta)$ is the normalized distribution of the surface brightness. Assuming that the cluster contribution to the XRB is fully described by Eq. (8), i.e., the total cluster contribution to the XRB is a simple sum of contributions of individual clusters, the CCF defined by Eq. (6) can be expressed in terms of a single cluster emission as follows:

$$w_{Xc}(\theta) = \frac{S_0}{\langle \rho_X \rangle} \left[ f(\theta) + n_c \int d^2\alpha \, w_{cc}(\alpha) f(|\boldsymbol{\theta} - \boldsymbol{\alpha}|) \right], \quad (9)$$

where $w_{cc}(\theta)$ is the cluster-cluster ACF and $n_c = \langle \rho_c \rangle$ denotes the average cluster density. In calculations leading to Eq. (9) we have expressed the cluster distribution $\rho_c(\mathbf{n})$ as a sum of delta functions and substituted it into Eq. (6). To interpret the two terms on the right hand side of Eq. (7), it is useful to use another definition of the CCF:

$$w_{Xc}(\theta) = \frac{\langle \rho_X(\theta) \mid c \rangle}{\langle \rho_X \rangle} - 1. \quad (10)$$

Here, the term $\langle \rho_X(\theta) | c \rangle$ denotes the average X-ray intensity at separation $\theta$ from a randomly chosen cluster. Accordingly, the first term in Eq. (9) describes the enhancement of the XRB at a distance $\theta$ from the cluster produced by this cluster and the second term gives the integrated contribution of other clusters which gather around. The cluster-cluster ACF, $w_{cc}$, describes the number of these clusters above the average cluster density. For a Poissonian cluster distribution ($w_{cc} \equiv 0$) the X-ray – cluster CCF is described by a single cluster profile of the X-ray surface brightness, but for the clumped cluster distribution, the CCF depends also on the shape of the cluster ACF.

In the first attempt to fit the model CCF described by Eq. (9) to the data we have assumed a standard "$\beta$"-model of the cluster emission:

$$f(\theta) = \frac{1}{2\pi\theta_0^2} \left[ 1 + (\theta/\theta_0)^2 \right]^{-3/2}, \quad (11)$$

where $\theta_0$ is the core radius and we have already put $\beta = 2/3$ (Jones & Forman 1984). We have calculated the cluster ACF in the three samples separately[1]. Because of the small number of objects involved in the calculations, the results are subject to large statistical noise. But in all three distance groups positive amplitude of the ACF is clearly present. The cluster ACF in the sample I is shown in Fig. 6 to illustrate the order of magnitude involved in the problem. Such strong clustering of objects on various scales plays a crucial role in our investigation.

The present model has 2 free parameters: the flux $S_0$ and the characteristic radius $\theta_0$. These two parameters have been fitted to the observed CCF in each sample separately. $S_0$ and $\theta_0$ were estimated by minimizing "$\chi^2$". The observational errors were approximated by the rms scatter between simulations. All the relevant parameters are listed in Table 1 and the "best fits" are shown in Figs. 5, 7 and 8. In the plots, points represent the actual data and crosses the model. Formal $1\sigma$ uncertainties calculated assuming $\Delta\chi^2 = 1$ (Avni, 1976) amount to around 10 % for all the fitted parameters.

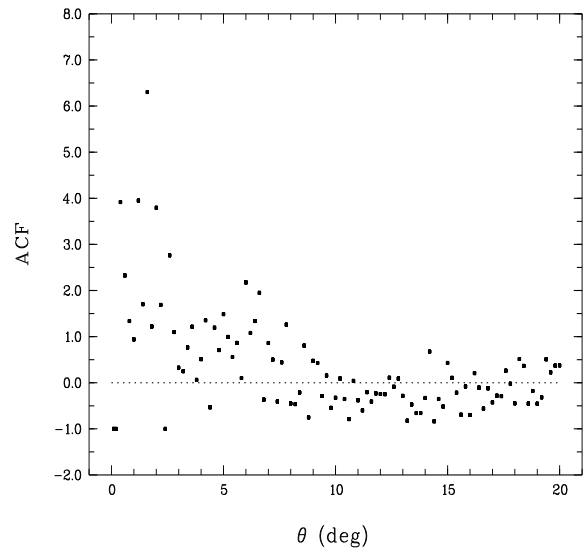

**Fig. 6.** The ACF of clusters in the sample I. A non-random distribution of objects at scales up to $\sim 8°$ is clearly visible.

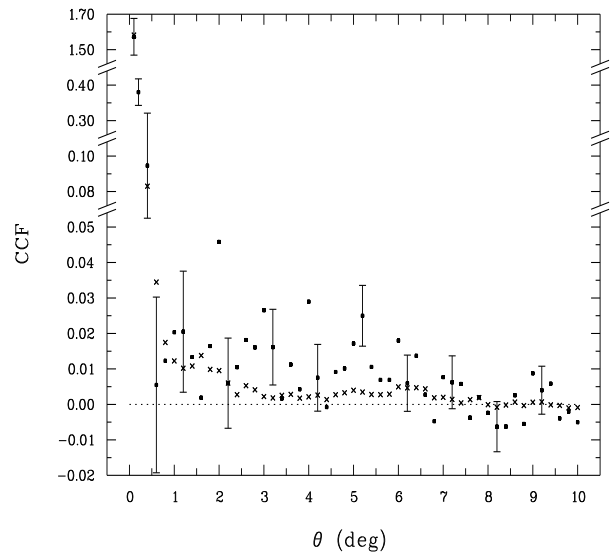

**Fig. 7.** The CCF of the XRB and clusters in sample I – dots, and the single component model CCF – crosses. The model adequately describes the strong X-ray emission at scales below $0°.5$ but fails to produce correlations at larger angular scales.

Before we discuss our results in detail we would like to state clearly the statistical limitations of this investi-gation. We have noted above that simulations reproduce all systematic effects and therefore provide authentic estimates of the total uncertainties involved in this study. Nevertheless, the present technique is not capable providing precise error assessments required for a scrupulous statistical analysis. In particular, accurate CCF errors are necessary to take full advantage of the $\chi^2$ statistics. The accuracy of our error estimates ascribed to a single CCF data point varies slightly with $\theta$, but we assess that errors are known typically with a precision of several percent. This implies that without any loss of accuracy it is legitimate to determine best values of free parameters by minimizing $\chi^2$, but we cannot use absolute $\chi^2$-values to eliminate/validate any particular model. Also, our estimates of uncertainty ranges of fitted parameters are accurate to several percent since they are based on $\chi^2$ increments.

---

[1] The ACF of the Abell clusters is fairly well known. However, in the present investigation we use specific subsamples of the Abell catalogue and we are interested in the actual cluster distribution within our data rather than the average correlations determined from the entire cluster population.

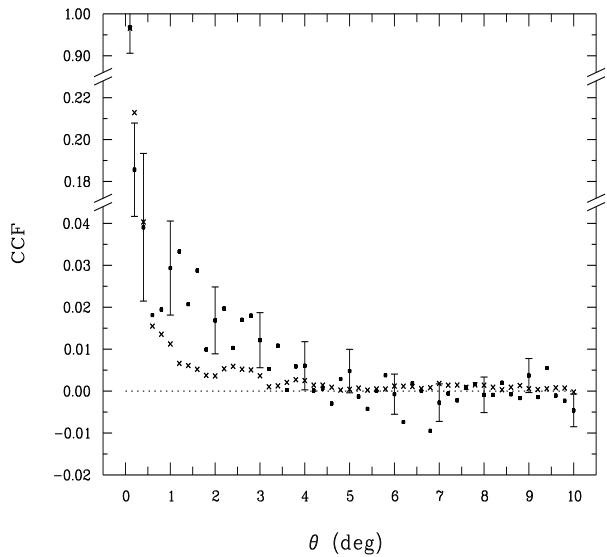

**Fig. 8.** Same as Fig. 7 for sample II.

Fluxes within the $0.5 - 2.0$ keV band in the second row of Table 1 have been calculated assuming that the cluster emission follows a thermal bremsstrahlung with $kT = 5$ keV (see below). Average redshifts $\langle z \rangle$ for each sample are given in the fourth row. Luminosities $L(0.5 - 2.0 \text{ keV})$ in the fifth row correspond to $S_0$ at those redshifts. A standard Friedmann model with a deceleration parameter $q_0 = 0$ and a Hubble constant $H_0 = 100 \cdot h \text{ km s}^{-1}\text{Mpc}^{-1}$ is used throughout the paper. Although the parameters $S$ and $L$ listed in Table 1 represent mean values obtained with drastic simplifications, they compare satisfactorily with results based on detailed cluster observations. In particular, the present estimates of the average cluster luminosities $L$ fit well to the recent measurements of the cluster luminosity function by Briel & Henry (1993). Their richness - luminosity relationship explains most of the variations of the average luminosity in our samples. The lowest luminosity found in sample I results from a high (57 %) fraction of the intrinsically least luminous RC = 0 clusters, and the luminosity difference between sample II and III is due to changing proportions of RC = 2 clusters in these samples. The combined contribution of all the Abell clusters in our three samples amounts to approximately 1.8 % of the total count rate in the R6 band.

Our estimates of the core radius in the three samples are subject to large systematic errors. First, there is a natural limit to measure small angular sizes caused by the pixel size of $12'$. However, due to the statistical character of our analysis we can effectively estimate core radii also below $12'$. This is because, even for $\theta_0$ substantially smaller than the pixel size, a measurable flux is detected not only in the pixel centered on the cluster but also in a few neighbouring pixels. The angular resolution of the off-axis PSPC data contributing a large fraction of the all-sky survey photons is of the order of few arc min what makes measurements of smaller diameters not feasible. Assuming a core radius of $a_0 = 0.125 \, h^{-1}$ Mpc (Henry et al. 1992) and adopting the average redshifts from Table 1, we expect $2'.5$, $1'.3$ and $0'.95$ for $\theta_0$ in sample I, II and II, respectively. This is much smaller than our measurements and demonstrates the significance of systematic effects.

The peak at small separations is reasonably well reproduced in our model for all three samples. However, at larger separations the results are more ambiguous. In all samples the model signal at separations above $\sim 0°.5$ is generated by the positive amplitude of the cluster-cluster ACF. This is because the second term in Eq.(9) becomes dominant on large angular scales. The values of $\theta_0$ are defined essentially by the width of the central peaks in the observed CCFs and the best fit $\theta_0$ is well below the pixel size in all the samples (Table 1). Thus, single cluster emission in the present model is localized and contributes practically only to the first three points of the CCF in Figs. 5, 7 and 8, i.e., for $\theta < 0°.5$, while the CCF amplitude on larger scales is generated by the non-uniform distribution of objects.

In the sample I (Fig. 7) the model CCF lies systematically below the observations. Apparently the large positive signal of the CCF on scales of several degrees cannot be generated by small angular size objects. A similar conclusion applies to the sample II (Fig. 8). Evidently, the model fails to reproduce correlations present in the data. The results for the sample III (Fig. 5) are non-restrictive. The CCF has a narrow peak at separations below $0°.3$ and this is the only statistically significant feature. A weak negative signal of $\approx -2 \times 10^{-3}$ visible in the data between $1°$ and $6°$ is statistically insignificant because the rms estimated from simulations exceeds $3 \times 10^{-3}$. The positive amplitude in the model CCF at $\sim 1°$ is not in conflict with the data due to the relatively large uncertainties involved in this problem.

One should expect that the substantial simplifications involved in our model prevent from precise measurements of source parameters. This results mostly from the fact that individual clusters produce distinctly different signals in the CCF. Estimates based on the cluster X-ray luminosity function (Briel & Henry 1993) show that only a small fraction of distant clusters produce count rates above the statistical noise in the data. Thus, the contribution to the observed $w_{Xc}(\theta)$ comes mostly from a few strong X-ray sources, while in the model calculations (Eq. 9) the cluster ACF is determined using all the clusters in the sample. Although this effect does not alter the expected values of the average cluster emission, it could introduce large errors. The problem concerns all three cluster samples but is particularly severe for the most distant objects. Consequently, in the following investigation we concentrate on the nearby clusters (sample I) and use medium distant clusters (sample II) as supplementary data. Because the sample III provides the least reliable results it will not be utilized in the subsequent analysis.

**Table 1.** One-component model parameters

|  | Sample | | |
|---|---|---|---|
|  | I | II | III |
| $S_0$ [PSPC cnt s$^{-1}$] | 0.078 | 0.036 | 0.019 |
| $S(0.5 - 2.0\,\mathrm{keV})$ [erg cm$^{-2}$s$^{-1}$] | $2.5 \times 10^{-12}$ | $1.2 \times 10^{-12}$ | $6.1 \times 10^{-13}$ |
| $\theta_0$ | $7'.1$ | $5'.4$ | $3'.1$ |
| $\langle z \rangle$ | 0.063 | 0.128 | 0.198 |
| $L(0.5 - 2.0\,\mathrm{keV})$ [erg s$^{-1}$] | $1.1 \times 10^{43} h^{-2}$ | $2.3 \times 10^{43} h^{-2}$ | $3.1 \times 10^{43} h^{-2}$ |

The two parameter model of the cluster emission defined by Eqs. (8) and (11) is not capable of producing large angular scale correlations between the XRB and clusters in samples I and II. This is because the CCFs exhibit a characteristic two component structure. The high amplitude, narrow peak is generated by standard, small angular size sources associated with clusters. To account for the second, low amplitude extended CCF feature, one needs an additional, diffuse component. Accordingly, we modify the model cluster emission as follows:

$$S(\theta) = S_{0c} \cdot f_c(\theta) + S_{0e} \cdot f_e(\theta), \tag{12}$$

where subscripts "$c$" and "$e$" denote the compact and extended components. The large uncertainties of the CCF measurements do not allow to investigate details of the surface brightness distribution of the extended component and it is assumed that both components are adequately described by formula (11) with the core radii $\theta_{0c}$ and $\theta_{0e}$.

The model CCF with 4 free parameters has been fitted to the observed CCF by minimizing $\chi^2$. Results for the sample I and II are compared in Table 2 and shown in Figs. 9 and 10. $S_{0c}$ and $\theta_{0c}$ of the compact component are practically the same as in the single component model because the parameters of both components are very weakly dependent. In particular, the CCF signal in sample I at large separations is reproduced by extended emission with a core radius $\theta_{0e}$ 31 times larger than the core radius of the compact emission. In the sample II this ratio is equal to 23. This demonstrates that the two distinct components are not artificially introduced by the present model but that they describe completely different X-ray properties of clusters. Linear sizes and luminosities of the extended sources ($a_e$ and $L_e$ in Table 2) are calculated using the average redshift for each sample. We estimate that the extended components around all Abell clusters account for about 2.6% of the total XRB flux. The uncertainty limits listed in Table 2 correspond to a confidence level of 68% assuming one interesting parameter (e.g., Avni, 1976) and do not include errors introduced by deficiencies of the model.

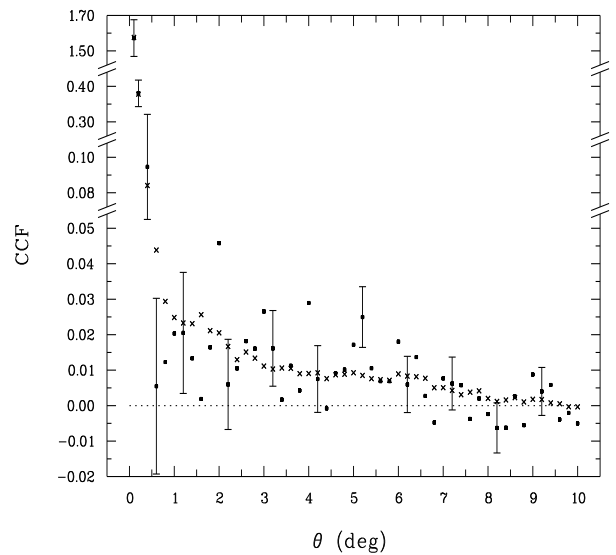

**Fig. 9.** The CCF of the XRB and clusters in sample I – dots, and the two component model CCF – crosses. The small angular size component reproduces the narrow peak of the CCF below $0°.5$, while the extended component generates a positive CCF amplitude up to $8°$.

**Table 2.** Two-component model parameters

|  | Sample | |
|---|---|---|
|  | I | II |
| Compact components | | |
| $S_{0c}$ [PSPC cnt s$^{-1}$] | 0.072 | 0.032 |
| $S_c(0.5 - 2.0\,\text{keV})$ [erg cm$^{-2}$s$^{-1}$] | $2.3 \times 10^{-12}$ | $1.0 \times 10^{-12}$ |
| $\theta_{0c}$ | $6'.6$ | $4'.8$ |
| Extended components | | |
| $S_{0e}$ [PSPC cnt s$^{-1}$] | 0.17 | 0.054 |
| $S_e(0.5 - 2.0\,\text{keV})$ [erg cm$^{-2}$s$^{-1}$] | $5.6^{+2.7}_{-1.9} \times 10^{-12}$ | $(1.8 \pm 0.4) \times 10^{-12}$ |
| $\theta_{0e}$ | $3°.4^{+1°.6}_{-1°.0}$ | $1°.3 \pm 0°.3$ |
| $a_e$ [Mpc] | $10.2^{+4.8}_{-3.0} h^{-1}$ | $7.3^{+1.8}_{-1.9} h^{-1}$ |
| $L_e(0.5 - 2.0\,\text{keV})$ [erg s$^{-1}$] | $2.5^{+1.2}_{-0.9} \times 10^{43} h^{-2}$ | $3.6^{+0.7}_{-0.8} \times 10^{43} h^{-2}$ |

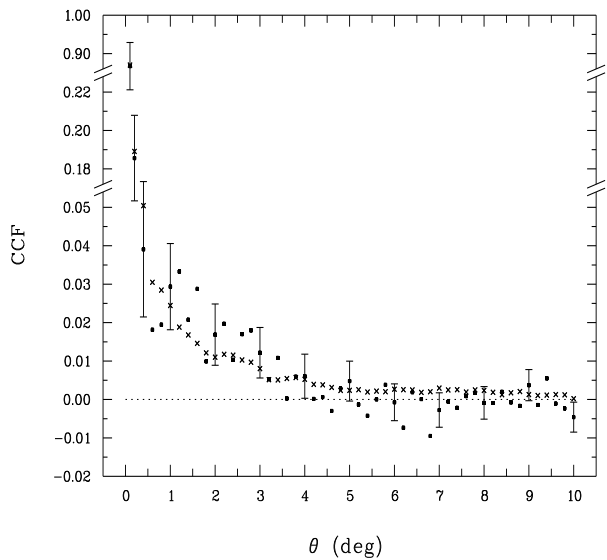

**Fig. 10.** Same as Fig. 9 for sample II.

Both visual impression and the $\chi^2$ statistics indicate that the two component model provides a substantially better representation of the data. This last conclusion is based on the $\chi^2$ variations rather than on the absolute amplitude of $\chi^2$ (see above). Although both models in sample I formally give quite satisfactory fits ($\chi^2$ are 78.23 (74 dof) and 60.27 (72 dof) for the one and two component model, respectively), we note that $\chi^2$ is substantially reduced for the latter model.

We would like to stress that the surface brightness, $S_e(\theta)$ of those sources is very low. The maximum, central $S_e(\theta = 0) = 6.5 \times 10^{-7}$ PSPC cnt s$^{-1}$arc min$^{-2}$, which is below 1 % of the total average XRB.

## 5. Model for extended X-ray emission around clusters

To convert the observed count rates of the extended sources to fluxes, we assume that the spectral characteristics of these sources are similar to the average extragalactic XRB. At energies relevant to our considerations this is well approximated by a power law with an energy spectral index of $-1$ (Hasinger 1992). The mean hydrogen column density in our data is close to $1.8 \times 10^{20}$ cm$^{-2}$. Using the standard calibration of the PSPC, we find for these parameters that a source generating count rate of 1 PSPC cnt s$^{-1}$ in the R6 band produces a flux outside the Galaxy of $3.29 \times 10^{-11}$ erg cm$^{-2}$s$^{-1}$ in the $0.5 - 2.0$ keV energy band. We note that a thermal bremsstrahlung source with $kT = 5$ keV (the average temperature for the standard cluster X-ray emission) would have the flux of $3.20 \times 10^{-11}$ erg cm$^{-2}$s$^{-1}$ to give the same count rate. The PSPC count rate in the band R6 varies slowly with temperature for a fixed integrated flux between 0.5 and 2.0 keV and for $kT = 1$ keV a flux of $3.17 \times 10^{-11}$ erg cm$^{-2}$s$^{-1}$ is needed to get 1 PSPC cnt s$^{-1}$ in the band R6.

We consider three possible mechanisms for the extended emission associated with clusters. First, we examine whether the enhanced density of galaxies and AGNs around Abell clusters could produce the diffuse emission detected here as extended sources. Next, we discuss a possibility that these apparent extended sources consist of discrete sources associated with various kinds of galaxy groups accumulated around rich Abell clusters. Finally, a truly diffuse model of thermal bremsstrahlung by low density gaseous haloes is investigated.

## 5.1. Enhanced galaxy density model

Analyzing the cross-correlation of the Abell clusters and Shane-Wirtanen galaxy counts, Seldner & Peebles (1977) found that the spatial density of galaxies around each cluster varies with the distance from the cluster centre. For clusters with the RC = 1 the mean density of galaxies is $n(r) \approx 165 \langle n \rangle (h\,r)^{-2.4}$ at separations $0.5\,\mathrm{Mpc} < h\,r < 15\,\mathrm{Mpc}$, where $\langle n \rangle$ is the average galaxy density. Although Seldner & Peebles note that there is some dependence of the galaxy density on the RC, we shall use the above formula for all the clusters in our samples. Substantial uncertainties involved in the present analysis do not warrant more accurate treatment. We now assume that the average X-ray volume emissivity $j_X(r)$ is proportional to the spatial galaxy density $n(r)$:

$$j_X(r) = [X/O]\, n(r), \qquad (13)$$

where $[X/O]$ denotes the overall ratio of X-ray–to–optical emissivities. We are interested only in the general correlations between the galaxy density and the X-ray emissivity. Although, one could expect some deviations from this formula in regions surrounding clusters, these would not affect the present consideration seriously. Using Eq. (13) and rewriting the Seldner & Peebles results in the form:

$$n(r) = \langle n \rangle \left( \frac{r}{r_0} \right)^{-2.4}, \qquad (14)$$

where $r_0 = 8.4 h^{-1}\,\mathrm{Mpc}$, we get the X-ray emissivity around clusters in terms the overall mean emissivity $\langle j_X \rangle$:

$$j_X(r) = \langle j_X \rangle \left( \frac{r}{r_o} \right)^{-2.4}, \qquad (15)$$

and integrated luminosity within a sphere of radius $r$:

$$L(r) \approx 21 r_0^3 \langle j_X \rangle \left( \frac{r}{r_0} \right)^{0.6}. \qquad (16)$$

The weak $L(r)$ dependence on $r$ allows us to make a comparison between our model of the extended source described in the previous section and the present luminosity estimates. Assuming that the total luminosity associated with the excess number of galaxies within $h\,r = 15\,\mathrm{Mpc}$ of the cluster corresponds roughly to the luminosity of the extended source in the sample I we get the mean X-ray volume emissivity:

$$\langle j_X \rangle = 1.4^{+0.7}_{-0.5} \times 10^{39} h\,\mathrm{erg\,s^{-1} Mpc^{-3}}. \qquad (17)$$

We compare this figure with the volume emissivity needed to reproduce the observed intensity of extragalactic XRB without cosmic evolution. According to Hasinger et al. (1993) a total extragalactic flux in the 1–2 keV band amounts to $1.25 \times 10^{-8}\,\mathrm{erg\,s^{-1}sr^{-1}}$. This gives $2.61 \times 10^{-8}\,\mathrm{erg\,s^{-1}sr^{-1}}$ in the 0.5–2 keV band assuming slope of the XRB spectrum $\alpha = -1.12$ (Hasinger 1992). The relationship between the volume emissivity $\langle j_X \rangle$ and the XRB flux $I_X$ for nonevolving sources with power law spectra takes the form:

$$I_X = \frac{1}{4\pi} \frac{c}{H_\circ} \langle j_X \rangle \int_0^{z_{max}} \frac{(1+z)^{1-\alpha}}{(1+z)^3 \sqrt{1+2q_\circ z}} dz, \qquad (18)$$

where $z_{max}$ is the maximum redshift for sources producing the XRB. Since there are no indications of any high redshift cutoff in the distribution of X-ray sources we put $z_{max} = \infty$ and note that results of the subsequent calculations for $z_{max} = 4$ differ by less than 1.5 %. According to Eq. (18), the total volume emissivity needed to produce the XRB is $j_X^{tot} = 2.73 \times 10^{39}\,h\,\mathrm{erg\,s^{-1} Mpc^{-3}}$. So the volume emissivity implied by enhancements of the XRB around Abell clusters exceeds 50 % of $j_X^{tot}$. This is slightly above the best estimates of the contribution of all sources associated with nearby galaxies. Lahav et al. (1993) find that all X-ray sources distributed locally as the general galaxy population account for $(30 \pm 15)\,\%$ of the XRB. This result compares favorably with $30^{+12}_{-6}\,\%$ obtained by Elvis et al. (1984) for the total contribution of AGN integrated over a wide range of luminosities. Both the Elvis et al. and Lahav et al. investigations refer to the 2–10 keV energy band of the *HEAO* A-2 instrument. Due to large uncertainties involved in the present calculations and in the Lahav et al. and Elvis et al. papers, all these results seem to be consistent although any detailed comparison between 0.5–2.0 keV band with the high energy data requires information on the average spectra of objects contributing to the XRB in a wide energy range.

Despite this apparent agreement between Eq. (17) and some (uncertain) numbers quoted from the literature on the X-ray volume emissivity, it is highly improbable that a larger density of galaxies produces X-ray haloes around rich clusters of galaxies. In the Sect. 5.4 below we calculate the predicted ACF of the XRB assuming that the Abell clusters of galaxies are the only source of fluctuations. Both compact and extended components as determined in the Sect. 4.4 are used. The ACF produced solely by clusters is shown in Fig. 2c with crosses. We note that the amplitude of this ACF varies from $\sim 4 \times 10^{-4}$ to $2 \times 10^{-4}$ for separations $2° - 5°$. Assuming that the extended cluster component is generated by X-ray sources associated with the overall galaxy population which contribute to the XRB in both energy bands (i.e., 0.5–2 keV and 2–10 keV), the expected ACF in the higher energy band should exhibit an amplitude similar to that at low energies. This prediction is in clear disagreement with upper limit of $1 \times 10^{-4}$ determined by Carrera et al. (1991, 1993) at energies 4–12 keV (solid line on Fig. 2d). We conclude that the extended component is not recorded by the *Ginga* experiment because it has a softer spectrum than the general XRB. This implies that haloes surrounding rich clusters are not dominated by contributions of individual sources associated with normal galaxies or AGNs.

### 5.2. Discrete source model

Clusters not included in the Abell catalogue potentially can introduce the large scale correlations between the Abell clusters and the XRB. The Abell catalogue contains rich and optically well defined clusters. However, less rich galaxy groups or clusters without a dense core are, as a class, also X-ray emitters. A non-uniform distribution of these objects around Abell clusters would produce a systematic average excess of the XRB discovered by means of the CCF.

First, we discuss the question whether a statistical incompleteness of the Abell catalogue could explain the apparent extended emission in samples I and II. Some clusters which satisfy all the selection criteria are missing from the Abell list. Assuming that the spatial distribution and X-ray properties of these objects are similar to the population of Abell clusters selected in the original list, one can estimate the number of these clusters required to produce the observed CCF amplitude. We applied a one-component model with the additional free parameter, viz. the average total cluster density, $n_c$. The best fit to the observed CCF using eq. (9) is obtained for $n_c$ larger than the actual Abell cluster concentration by a factor of $2.6^{+0.5}_{-0.4}$ in the sample I and of $2.8^{+0.6}_{-0.5}$ in the sample II. Although varying $n_c$ allows the reproduction of the CCF signal at large separations quite well, the implied number of missing "Abell" clusters from the catalogue is very high. Taking these figures at their face value would indicate that the Abell catalogue contains only roughly 35 - 40 % of all clusters satisfying the selection criteria. This seems rather unlikely and indicates that objects potentially producing the extended emission in samples I and II are statistically different from "ordinary" Abell clusters.

The statistical completeness of the Abell catalogue refers to rich clusters with more than 50 member galaxies in the magnitude interval of 2 mag below the third brightest cluster member (RC $\geq$ 1). All varieties of smaller galaxy agglomerations left outside the Abell list could potentially produce the extended haloes around Abell clusters. Our sample I contains all objects from the Abell list, including RC = 0 clusters which do not belong to a complete subset of the catalogue, while the sample II is restricted to RC > 0 clusters. We now tentatively assume that the extended CCF signal in both samples is produced by clusters satisfying RC = 0 counts which were omitted in the samples I and II. First, we calculate CCFs between the XRB and RC = 0 clusters for distant classes corresponding to our original samples I and II, i.e., for DC = 1 − 4 and DC = 5, respectively. These CCFs are used to calculate an average flux produced by RC = 0 clusters in the one-component model. For DC = 0 − 4, the average flux is $S_0 = 0.062\,\mathrm{PSPC\,cnt\,s^{-1}}$ and for DC = 5 $S_0 = 0.016\,\mathrm{PSPC\,cnt\,s^{-1}}$. The extended haloes around clusters in samples I and II give fluxes of 0.17 and $0.054\,\mathrm{PSPC\,cnt\,s^{-1}}$, respectively (Table 2). Thus, the average number of missing RC = 0 clusters needed to mimic a single extended halo around each cluster in the sample I is approximately 2.7, and in the sample II it is 3.3. In the northern galactic hemisphere RC = 0 Abell clusters amount to more than 50 % of all DC$\leq$ 4 clusters. The number of RC = 0 clusters decreases with distance and for DC = 5 these clusters constitute about one third of all clusters. Our results indicate that the space density of hypothetical clusters missing from the Abell catalogue in both distance groups exceeds several times the number of known rich clusters. The number becomes even larger if one assumes clusters and groups of even lower richness. To verify this conjecture one should investigate the distribution of discrete sources around Abell clusters and look for identifications of these sources with groups of galaxies.

### 5.3. Thermal emission model

The extended source producing the surface brightness distribution given by Eq. (11) has the volume emissivity:

$$j_X(r) = j_X(0)\left[1 + (r/a_e)^2\right]^{-2}, \qquad (19)$$

where the central emissivity is related to the total luminosity $L$ and the core radius $a_e$:

$$j_X(0) = \frac{L}{\pi^2 a_e^3}. \qquad (20)$$

In the subsequent calculation we set the plasma temperature to $kT = 1\,\mathrm{keV}$, but all the numerical results hold within 6 % also for $kT = 5\,\mathrm{keV}$. Assuming that the extended emission is produced by isothermal plasma and taking $L(0.5 - 2.0\,\mathrm{keV})$ and $a_e$ from the sample I we get the emissivity at 1 keV:

$$j_X(0) = 6.1^{+15.8}_{-4.4} \times 10^{-35} h \,\mathrm{erg\,s^{-1}cm^{-3}keV^{-1}}. \qquad (21)$$

To determine uncertainty limits, a 68 % confidence region for two interesting parameters, $L$ and $a_e$ has been determined (Avni 1976). The quoted errors correspond to the minimum and maximum values of $j_X(0)$ found for combination of $L$ and $a_e$ in that region. The spectral thermal emission per unit volume of hot plasma (Zombeck 1982):

$$\frac{dP}{dV\,d\nu} = 6.8 \times 10^{-38} T^{-1/2} \exp(-E/kT)$$
$$n_e n_Z Z^2 \overline{g(E,T)}\,\mathrm{erg\,s^{-1}cm^{-3}Hz^{-1}}, \qquad (22)$$

where $\overline{g(E,T)}$ is the Gaunt factor and the remaining symbols have their usual meaning. Using standard cosmic abundances (Morrison & McCammon 1983) and Gaunt factor approximation by Kellogg et al. (1975), Eqs. (21) and (22) provide an estimate of the central electron density:

$$n_e(0) = 4.2^{+3.8}_{-2.0} \times 10^{-6} \, h^{1/2} \, \text{cm}^{-3}, \qquad (23)$$

and the central gas density $\rho_g(0)$:

$$\rho_g(0) = 8.3^{+7.5}_{-4.0} \times 10^{-30} \, h^{1/2} \, \text{g cm}^{-3}. \qquad (24)$$

An isothermal gas cloud emitting according to Eq.(19) forms a configuration of divergent mass with increasing distance from the centre. We do not consider this as an important difficulty for our analysis because the present model is designed to describe only the central regions of highest surface brightness within a couple of core radii $a_e$. Given the substantial uncertainties it is not feasible to determine whether individual sources have well defined boundaries or join smoothly with other sources associated with neighbouring clusters. The mass of the plasma contained in a sphere of radius $a_e$ is equal:

$$\begin{aligned} M_g(a_e) &= 7.0^{+11.0}_{-4.3} \times 10^{47} \, h^{-2.5} \, \text{g} \\ &= 3.5^{+5.4}_{-2.1} \times 10^{14} \, h^{-2.5} \, M_\odot. \end{aligned} \qquad (25)$$

If the gas remains in the hydrostatic equilibrium with the gravitational potential, the total mass within $a_e = 10.2\,\text{Mpc}$ is $M(a_e) = 3.7 \times 10^{14}(kT/1\,\text{keV})\,M_\odot$. The hydrostatic equilibrium by no means is implied by our data and it is assumed here just to estimate orders of magnitudes of quantities involved in this investigation.

The mass of the gaseous haloes given by Eq. (25) is not in conflict with any assessment of the baryonic mass density in the Universe. For a standard Friedmann model with total density parameter $\Omega = 1$, the average density of matter collected in clumps of mass $M_g(a_e)$ in the volume occupied by the sample I amounts approximately to:

$$\overline{\rho_g} = 1.6 \times 10^{-31} \, h^{0.5} \, \text{g cm}^{-3}, \qquad (26)$$

and the corresponding density parameter $\Omega_g = 0.0087\,h^{-1.5}$.

Finally, we would like to estimate the amplitude of the SZ effect on the microwave background radiation generated by the gas clouds in the present model. The total optical depth of the extended source $\tau = \pi \sigma_T a_e n_e(0)$, where $\sigma_T$ is the Thompson cross-section, and for the "best estimates" of $a_e$ and $n_e(0)$ given above, $\tau = 2.7 \times 10^{-4} \, h^{-1/2}$. A relative amplitude of the temperature variation in the Reyleigh-Jeans region of the Planck spectrum is given by (Sunyaev & Zel'dovich 1980):

$$\frac{\Delta T_{rad}}{T_{rad}} = -\frac{2kT}{m_e c^2}\tau, \qquad (27)$$

which for $kT = 1\,\text{keV}$ gives $\Delta T_{rad}/T_{rad} = -1.1 \times 10^{-6} \, h^{-1/2}$. That is an order of magnitude below the rms variation of $11 \times 10^{-6}$ measured by COBE (Smoot et al. 1992). We note, that if higher gas temperatures are adopted, extended sources could produce signal in the microwave background comparable to the COBE detection.

### 5.4. Abell cluster contribution to the ACF of the X-ray background

Emission associated with the Abell clusters produces noticeable fluctuations of the XRB. Obviously, other classes of sources generate variations on their own, but Abell clusters are relatively strong X-ray sources which produce a measurable signal in the ACF of the XRB. The contribution of Abell cluster to the ACF has been assessed as follows. Using the average source parameters determined for the sample I, we have scaled luminosities and angular sizes of sources associated with the remaining, more distant clusters. Then a simulated distribution of the X-ray surface brightness was generated. Sources with given flux and size have been superimposed on the smooth count rate distribution at positions of the actual Abell clusters. Fluxes of compact components which represent standard cluster X-ray emission (Sect. 4.4) have been randomized according to the cluster X-ray luminosity function. The luminosity function of the extended components remains undetermined, thus the average luminosities were used in the calculations. The ACF of the simulated count rate distribution is shown with crosses in Fig. 2c. At separations $\theta > 4°$ the amplitudes of the observed and simulated ACF are similar. For smaller separations the signal expected from Abell clusters is roughly 3 times weaker than that actually detected. This implies that other sources produce still a large fraction of the observed fluctuations. It is highly probable that various smaller groups of galaxies could generate the remaining XRB variations. The problem of large scale fluctuations produced by these groups will be analysed in a future investigation.

The ACF amplitude of the XRB attributed to the Abell clusters exceeds $1 \times 10^{-4}$ for separations below $10°$ and rises above $2 \times 10^{-4}$ – below $5°$. To reconcile such strong signal with the GINGA results (Carrera et al. 1991, 1993) one has to assume that the cluster emission is much weaker in the GINGA energies of $4-12\,\text{keV}$ than in the ROSAT band. This is unquestionable in the case of the standard cluster emission, for which the mean temperature of $5.5\,\text{keV}$ has been obtained by David et al. (1993). We conclude that to satisfy the Ginga constraints, $\sim 5\,\text{keV}$ upper limit for the temperature of the extended component is required.

## 6. Concluding discussion

Large regions of enhanced X-ray emission surrounding Abell clusters constitute a new class of X-ray sources. It is shown that the extended emission is not supplied by normal galaxies which accumulate around rich clusters in sizable haloes.

We tentatively assume that the extended features related with the Abell clusters are produced either by a large number of discrete sources associated with groups of galaxies and poor clusters or by hot gas clouds with typical diameters of $\sim 20$ Mpc. In both cases the X-ray emission has to be softer than the overall XRB flux to avoid discrepancies with stringent upper limits of the background fluctuations given by Carrera et al. (1991, 1993).

Previous searches for extended emission using the *HEAO 1* A-2 data base (Persic et al. 1988, 1990) gave negative results. These authors looked for the enhanced X-ray flux from directions associated with superclusters. In the second paper Persic at al. gave a $3\sigma$ upper limit to the mean excess flux of $5 \times 10^{-12}$ erg s$^{-1}$ cm$^{-2}$ in the 2 - 10 keV band. They note, however, that the sensitivity of their measurements is reduced to $\sim 20\%$ for a source angular radius of $3°$ and to $10\%$ for a radius of $4°$. This implies that the extended sources reported in the present paper are below the detection threshold of the *HEAO 1* data.

The mean flux produced by the extended component associated with each cluster in sample I ($DC \leq 4$) $S_{0e} \approx 6 \times 10^{-12}$ erg cm$^{-2}$ s$^{-1}$. Due to the large angular size, the surface brightness of these sources is very low and the signal generated by a single source is below the detection threshold. Since the statistical analysis based on the CCF provides only information on the average properties of the sources in the whole sample, we are not able to identify any particular object. Nevertheless, direct inspection of X-ray and cluster maps reveals some striking correspondence of the large scale structure (superclusters and voids) in both distributions. This visual impression cannot be quantitatively confirmed for any individual feature but it is in fact verified by our correlation analysis.

The X-ray flux correlated with the Abell clusters constitutes $\sim 4\%$ of the total XRB in the R6 band. Sources associated with clusters generate complex variations of the XRB. Each source consists of the compact component, which represents standard cluster emission coinciding with the optical position of a cluster, and the extended component with a characteristic scale of $\sim 10$ Mpc. Due to the non-random distribution of clusters, the angular scale and amplitude of the XRB fluctuations produced by clusters depend on the average parameters describing a single source as well as on the clumpy nature of the cluster distribution. Therefore, both the ACF of the XRB, and the CCF of the XRB with Abell clusters are related to the ACF of the cluster sample. We show that the Abell clusters contribution to the total XRB fluctuations amounts to about 30%. The source of the remaining variations is not explained, though it is likely that various nearby groups of galaxies could account for the excess fluctuations.

Hogan (1992) investigated the potential role of the SZ effect for the microwave background anisotropy detected by the *COBE* DMR experiment. He considered hot diffuse clouds of gas created during the supercluster collapse with densities and sizes very similar to those found in the present paper. Hogan has shown that the background variations produced by those clouds may mimic the primordial fluctuations. The question of correlations of the RASS fluctuations with the *COBE* maps will be discussed in a subsequent paper.

*Acknowledgements.* The *ROSAT* project has been supported by the Bundesministerium für Forschung und Technologie (BMFT/DARA) and by the Max-Planck-Society. AMS is grateful to Prof. J. Trümper for his hospitality and financial support. This work has been partially supported by the Polish KBN grant 2 P304 021 06. We thank the referee, Professor A. C. Fabian, for his remarks which helped us to broaden our discussion on the cluster relationship to the XRB.